\documentclass[twocolumn,tighten]{aastex631}
\usepackage{graphicx}
\usepackage{appendix}
\usepackage{url}
\usepackage{hyperref}
\usepackage{amsmath}
\usepackage{footnote}

\usepackage{color}

\begin{document}


\title{\fontsize{11pt}{12pt}\selectfont 
Jet collimation in a spiral-hosted AGN: a parabolic jet profile in 0313--192
}

\author[0009-0005-2960-3088]{Seung Yeon Lee}
\affiliation{
Department of Physics, Ulsan National Institute of Science and Technology (UNIST), 50 UNIST-gil, Eonyang-eup, Ulju-gun, 44919 Ulsan, Republic of Korea
}

\author[0000-0001-8229-7183]{Jae-Young Kim}
\affiliation{
Department of Physics, Ulsan National Institute of Science and Technology (UNIST), 50 UNIST-gil, Eonyang-eup, Ulju-gun, 44919 Ulsan, Republic of Korea
}
\correspondingauthor{Jae-Young Kim}
\email{jaeyoungkim@unist.ac.kr}


\begin{abstract}

Double-lobed radio sources associated with active galactic nuclei (DRAGNs) are typically found in elliptical galaxies, while supermassive black holes (SMBHs) in disk galaxies rarely produce powerful kpc-scale jets. However, the growing number of spiral- and disk-hosted DRAGNs challenges this classical dichotomy. We present a study of the jet collimation profile for one such source, 0313--192, using VLBA and VLA data, tracing the jet morphology across nearly five orders of magnitude in scale---from $\sim$ pc to $\sim100$\,kpc (projected). We find that the jet exhibits a parabolic expansion up to $\sim 610$\,pc ($\sim 7.9 \times 10^6$ Schwarzschild radii), followed by a transition to a nearly conical shape, assuming kpc-scale emission primarily originates from the jet rather than the lobe. This structural evolution closely resembles those in AGNs hosted by elliptical galaxies and provides an explanation for how the jet in this system could extend to large distances by magnetohydrodynamic collimation and acceleration. However, this collimation break occurs beyond the sphere of gravitational influence of the SMBH ($\sim7.3\times10^{5}R_{S}$), and no extended X-ray halos or dense molecular gas structures are detected to provide the necessary external pressure. Therefore we suggest that jet confinement in 0313--192 is mediated by contributions from non-thermal components, such as ram and magnetic pressure from magnetized disk winds. These mechanisms may enable jet collimation even in the absence of dense ambient gas. Our results highlight how large-scale jets can arise in disk galaxies under rare conditions and demonstrate the need to broaden studies of AGN jet formation beyond traditional models.

\vspace{\baselineskip}

\end{abstract}


\vspace{2\baselineskip}

\section{Introduction} \label{sec:intro}

The physical origin of relativistic jets in active galactic nuclei (AGNs) remains a long-standing question in astrophysics (see \citealt{Blandford_2019} for a review). While extraction of rotational energy from spinning supermassive black holes (SMBHs) or their accretion disks is widely accepted as the primary jet-launching mechanism \citep{bz77,bp82}, powerful, well-collimated, and large-scale ($>$kpc) jets---so-called double-lobed radio AGNs (DRAGNs; \citealt{Leahy_1993})---are almost exclusively observed in giant elliptical galaxies or merging systems, often located at the centers of clusters \citep{Eilek}. This highlights the importance of host galaxy environments in determining the growth and properties of jets (e.g., \citealt{condon92,urry95,Ho_2008,Heckman_2014,kim18}).
In particular, hot, geometrically thick accretion flows (so-called advection-dominated accretion flows or ADAFs; \citealt{Ichimaru_1977,Narayan_1994,Narayan_1995}), commonly found in ellipticals, are believed to support jet collimation and acceleration to galactic scales (e.g., \citealt{komissarov07,Asada_2012,Boccardi_2017,kim19,Blandford_2019}).

However, recent deep radio and optical surveys have revealed a growing population of DRAGNs hosted by spiral or disk galaxies, often without signs of major mergers \citep[e.g.,][]{Hota2011,Bagchi2014,singh2015,Wu_2022,bagchi2024,sethi2025}. These so-called spiral DRAGNs challenge classical jet formation scenarios, as their AGNs are likely powered by geometrically thin, radiatively efficient disks \citep{shakura73,Heckman_2014}, which are not typically associated with strong jet production. Moreover, unlike ellipticals, spirals generally lack hot halo gas from large-scale cooling flows \citep{Nulsen_1984,Thomas_1986,Mathews_1999,Bregman_2001,Bregman_2004,Ivey_2024}, unless enhanced by recent interactions.
These findings motivate broader theoretical models involving thin accretion disks as important ingredient of the jet launching models,  especially those incorporating strong magnetic fields and disk winds (e.g., \citealt{Liska_2019,Dihingia_2021,Lalakos2024}; see also \citealt{mizuno22}).

To explore this further, we investigate the collimation profile of the large-scale jet in the spiral galaxy 0313--192 (WISE J031552.09--190644.2; 2MASS J03155211--1906442), one of the first identified spiral DRAGNs \citep{Ledlow_1998}. At a redshift of 0.067, 0313--192 shows $\sim350$\,kpc radio jets in the Very Large Array (VLA) images and an edge-on Sb morphology with a Seyfert nucleus in Hubble Space Telescope (HST) observations \citep{Ledlow_1998,ledlow2001large,Keel_2006}. Also its compact core is detected at 1.4--8.5\,GHz with the Very Long Baseline Array (VLBA) \citep{mao2018}, enabling very-long-baseline interferometry (VLBI) imaging of the inner pc-scale jet as well. 
Notably, its central SMBH is unusually massive for a spiral galaxy ($M_{\rm BH}\sim 8\times10^8 M_{\odot}$; \citealt{Keel_2006}). Combined with the distance scale of 1.3\,pc per milliarcsecond (mas), this allows high-resolution imaging of the object down to $\sim1.7\times10^4R_{S}$ resolution per mas and compare the results with those of  high-spatial-resolution studies of jet collimation profiles in nearby elliptical galaxies (e.g., \citealt{Asada_2012,Boccardi_2015,Tseng_2016,Nakahara_2018,Nakahara_2019,Boccardi_2021,Park_2021,Hada_2018,Boccardi_2019,Giovannini_2018,Okino_2022,Kovalev_2020}).

In this letter, we adopt a $\Lambda$CDM cosmology with $H_{0} = 71 \, \text{km s}^{-1} \text{Mpc}^{-1}$, $\Omega_M = 0.27$, and $\Omega_\Lambda = 0.73$, following \cite{Spergel_2003}, consistent with the angular-to-spatial scaling factors adopted in \cite{mao2018}.



\section{Data Reduction \& Analysis} \label{sec:data}

For this study, we used archival VLA observations at 1.4 and 8.5\,GHz (project codes AL400 and AY119 in A and CnD configurations; \citealt{ledlow95,Ledlow_1998}) and VLBA observations at 1.4, 2.3, and 8.5\,GHz (project code BM376; \citealt{mao2018}), retrieved from the National Radio Astronomy Observatory (NRAO) data archive\footnote{\url{https://data.nrao.edu/}}.
Table~\ref{tab:data} summarizes basic information of the observations. More details (such as observing hours and bandwidths) can be found in \cite{Ledlow_1998} and \cite{mao2018}.
In this study, we omit the analysis of the structure of the northern (presumably) counterjet of 0313--192, seen on the arcsecond scales \citep{ledlow2001large} but not on the milliarcsecond scales by the VLBA \citep{mao2018}.

Data calibration was performed using the Astronomical Image Processing System (AIPS; \citealt{greisen03}). 
For the VLA data, standard procedures were used to determine the complex antenna gains for the main source from calibrators (see also \citealt{ledlow95}).
For the VLBA data, special care was taken to flag radio frequency interference (RFI)-affected channels, particularly at the S-band (2.3\,GHz). Otherwise, standard calibration steps followed, including corrections for ionospheric delays, bandpasses, instrumental delays and phases, and amplitudes. Since the data were taken in phase-referencing mode, we tested both interpolated solutions from calibrator source J0315--1656 and direct fringe fitting of 0313--192. 
Both approaches provided reliable fringe detections and comparable fringe solutions at all frequency bands. 
Since the direct fringe fitting yielded solutions with higher time resolution, we adopted those solutions for later analysis.
Imaging was performed in the Difmap software \citep{shepherd97} using various $(u,v)$ weighting and tapering schemes, to optimize sensitivity to both compact and extended structures.
The $(u,v)$-tapering was applied by using a Gaussian-shaped tapering function that down-weights long-baseline visibilities. This is implemented in Difmap via the \texttt{uvtaper} command, where the strength and spatial extent of the tapering is controlled by specific amplitude of the weight at a given $(u,v)$-radius.
Combination of various weights and tapering led to maps with various angular scales and sensitivities to the faint and extended jet. 
Each imaging run was performed by iterative CLEANing and self-calibration, with pixel sizes set to 1/5 of the synthesized beam, considering the resolution limit \citep{lobanov2005}.
The adopted weighting and tapering schemes are summarized in Table \ref{tab:data} of Appendix~\ref{sec:more_details}. 

In order to measure the transverse jet width as a function of distances from the peak of the image, 
we performed Gaussian fitting to the jet-transverse intensity profiles and computed beam-deconvolved Gaussian full-width at half-maximum (FWHM) jet widths. 
Following \cite{Pushkarev_2017}, VLBA maps were stacked per frequency and weighting scheme to improve the signal-to-noise (S/N) and trace the time-averaged jet structure.
Each stacked image was restored with a circular beam and sliced perpendicular to the jet axis in steps of one-fifth the beam size, considering the effective resolution limit \citep{lobanov2005,Tseng_2016}. The slicing started from one beam size away from the peak of each map, to avoid regions where the jet emission is significantly blended by the core. Cross-jet intensity profiles were fitted with Gaussians if the peak exceeded $>\times3$ of the image rms noise, and beam-deconvolved jet FWHM widths, $d$, were computed as $d = \sqrt{W^2 - B^2}$, where $W$ is the measured width and $B$ the beam FWHM. 

Physically speaking, we consider that emission seen in the VLBA-scale images represent relativistic jet, given the absence of counterjet in the images and accordingly relatively high underlying speed of the emitting plasma \citep{mao2018}.
However, on large arcsecond scales, especially in the VLA images, the presence of large-scale emission in the north of the core suggests potential lobe origin of the emission.
Indeed, slice intensity profiles sometimes showed signatures of not only the jet but also additional broad and diffuse component, which could be associated with the lobe.
To accurately measure the jet width, especially for the VLA images, we decomposed the intensity profiles into the jet and lobe through double Gaussian fitting, similar to analyses performed by \cite{Nakahara_2018,Nakahara_2019} for large-scale Fanaroff-Riley (FR) type I and II jets. After each fit, we identified a narrower, high-peaked Gaussian component as the jet and associated the other broad component as lobe. 
Otherwise, if a single Gaussian was sufficient to describe the jet-transverse intensity profile (especially for the VLA X-band A configuration image), we took its FWHM as the jet width. 
Thus we assume that the kpc-scale emitting structure primarily consists of the jet and not the lobe.
See Fig.~\ref{fig:XCnD_slicing} in Appendix \ref{sec:more_details} for an example of the procedure.
In the above procedures, we also manually inspected the fit quality by checking large fit residuals or clear outliers and applied manual flagging. 

To construct the global jet collimation profile, we compiled width measurements across all maps in different frequencies as a function of core distance. In doing so we assumed that all the multi-frequency images could be aligned by their peak positions, which we defined as the jet ``core''. This was motivated by the fact that in all radio images of 0313--192 the jet morphology is simple and the apparent origin of the jet coincides with the peak points at all frequencies and resolutions.
We note that this alignment should bring some systematic uncertainties due to the coreshift effect \citep{lobanov1998}. The coreshift effect could introduce up to $\sim1\,$mas image misalignment at the $\sim$2--8\,GHz frequency range \citep{kovalev08,sokolovsky11,plavin19}.
This amount could affect calculation of the core distance for the jet width measurement at X-band (8\,GHz) where a few mas-long jet is transversely resolved, and thus additional care may be necessary.  However, in other frequency maps we start to resolve the jet only at much larger $\gtrsim10$\,mas distances from the intensity peaks (see Sect. \ref{sec:results}). Thus those width measurements should not be sensitive to the aforementioned coreshift effect (see also \citealt{kovalev20}).

We then modeled the global jet width profile using a broken power-law function \citep{Boccardi_2021}:
\begin{equation}\label{eq:jetwidth}
d(z) = d_t  2^{(k_u-k_d)/h}  \left(\frac{z}{z_t}\right)^{k_u}  \left[1+\left(\frac{z}{z_t}\right)^h\right]^{(k_d-k_u)/h}
\end{equation}
\noindent where
$d(z)$ is the jet width at the core distance $z$, 
$d_{t}$ the width of the jet at a transition point distance $z_{t}$ where the transition of the power-law of the collimation profile occurs,
$k_{u}$ and $k_{d}$ the slopes of the power-law of the jet width before and after the transition point, respectively, and
$h$ a factor describing the smoothness of the power-law transition at $z_{t}$.
To ensure that only independent jet width measurements were used in fitting Eq. \ref{eq:jetwidth}, we retained a single value of $d(z)$ at each core distance multiple measurements existed within a certain range for a given frequency, keeping the width from the tapered image due to its higher sensitivity to the broader jet structure.
This selection was applied primarily to the VLBA L-band measurements.

%


\section{Results} \label{sec:results}


\begin{figure*}[t!]
\centering
\resizebox{\textwidth}{!}{
\plotone{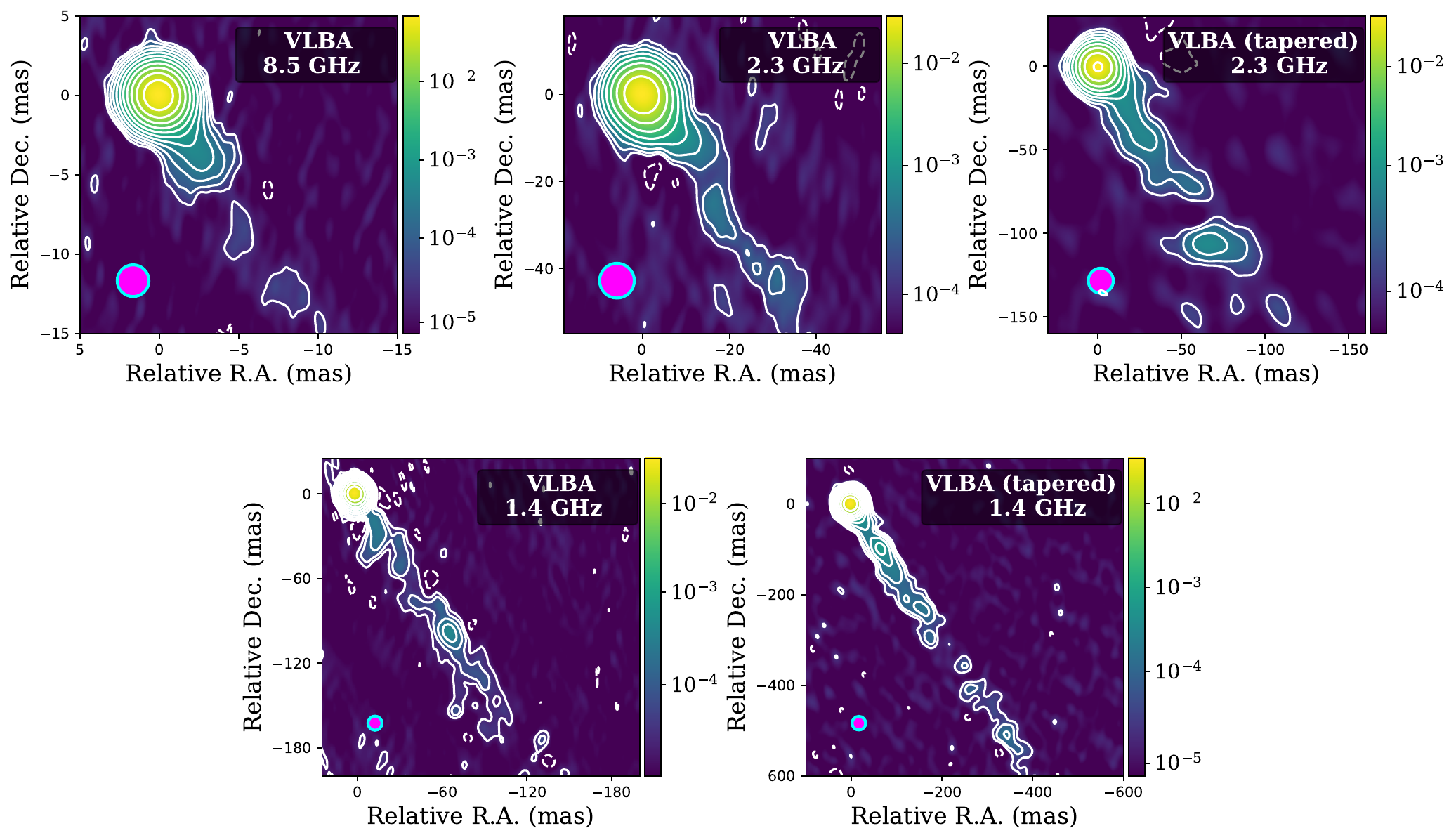}
}
\vspace{12pt} 
\resizebox{\textwidth}{!}{
\plotone{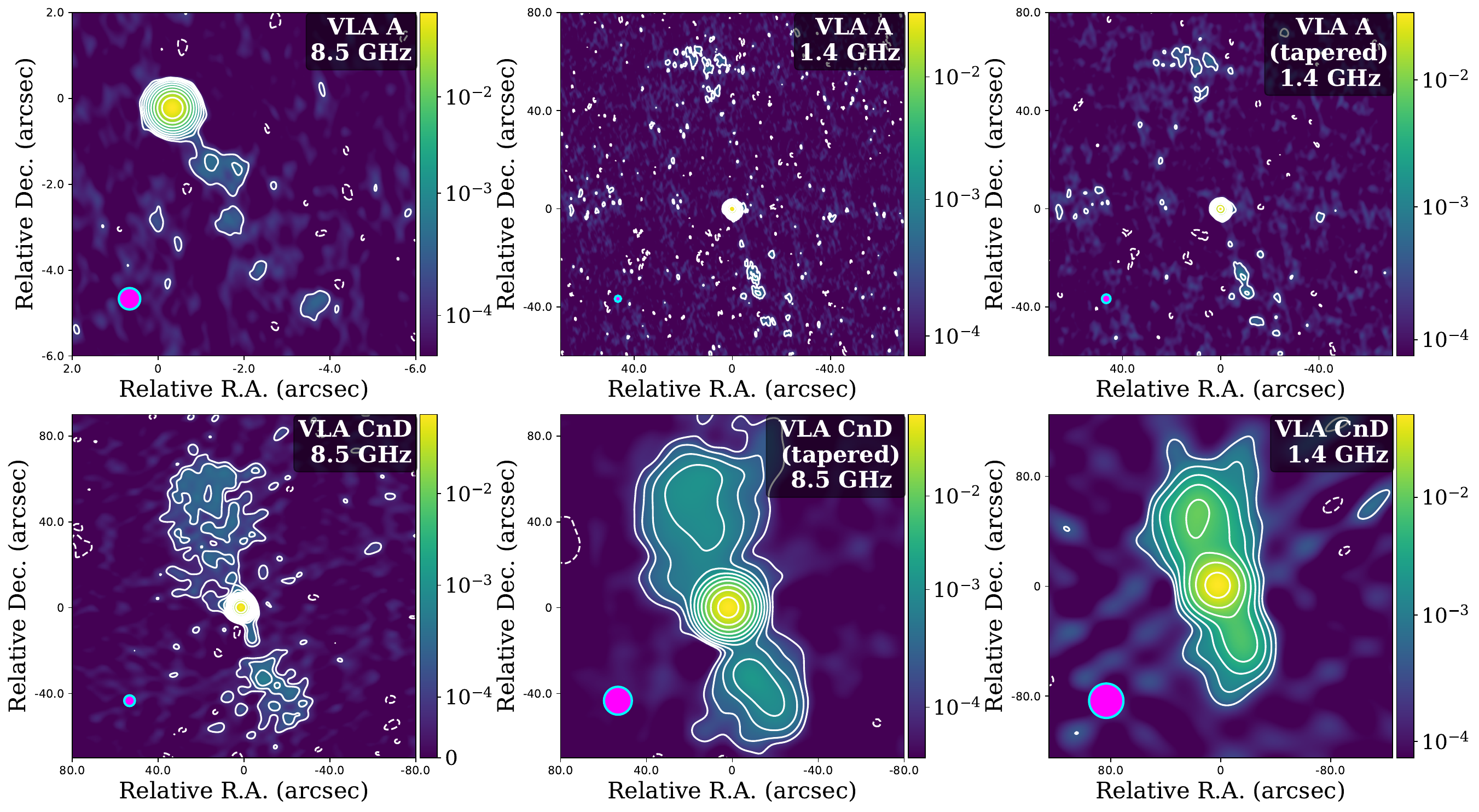}
}
%
%
\caption{
Stacked VLBA and VLA images of 0313--192. Upper panels: Images for VLBA X-band (top-left), S-band (top-center), tapered S-band (top-right), L-band (bottom-left), and tapered L-band (bottom-right). Restoring beam sizes are 2, 8, 15, 10, and 30 mas, respectively. Lower panels: VLA images in X-band A (top-left), L-band A (top-center), tapered L-band A (top-right), X-band CnD (bottom-left), tapered X-band CnD (bottom center), and L-band CnD (bottom-right) configurations. Restoring beam sizes are 500, 2500, 3500, 5000, 13000, and 25000 mas, respectively. Contours in all images start at 2.5\,$\sigma$ image noise levels and increase by factors of 2. The color scale indicates total intensity in Jy/beam.
\label{fig:imshow-all}
}
\end{figure*}

The individual VLBA images of 0313–192 across multiple bands and epochs are shown in Fig.~\ref{fig:XSL-imaging1} of Appendix~\ref{sec:more_details}. Also, the time-averaged, stacked VLBA and single-epoch VLA images are presented in Fig.~\ref{fig:imshow-all}. 
While the individual images have comparable morphologies and sensitivities as those shown in \cite{ledlow2001large} and \cite{mao2018}, the tapered and stacked images better reveal the diffuse jet emission.
Additionally, we note that the jet emission in the VLA L-band A-configuration image is heavily resolved due to the high resolution, compared to the VLA CnD configuration images that probe similar field of view. Therefore, we additionally excluded the VLA L-band A-configuration measurements in the collimation profile fitting.

\begin{figure*}[t!]
\centering
\resizebox{\textwidth}
{!}{\plotone{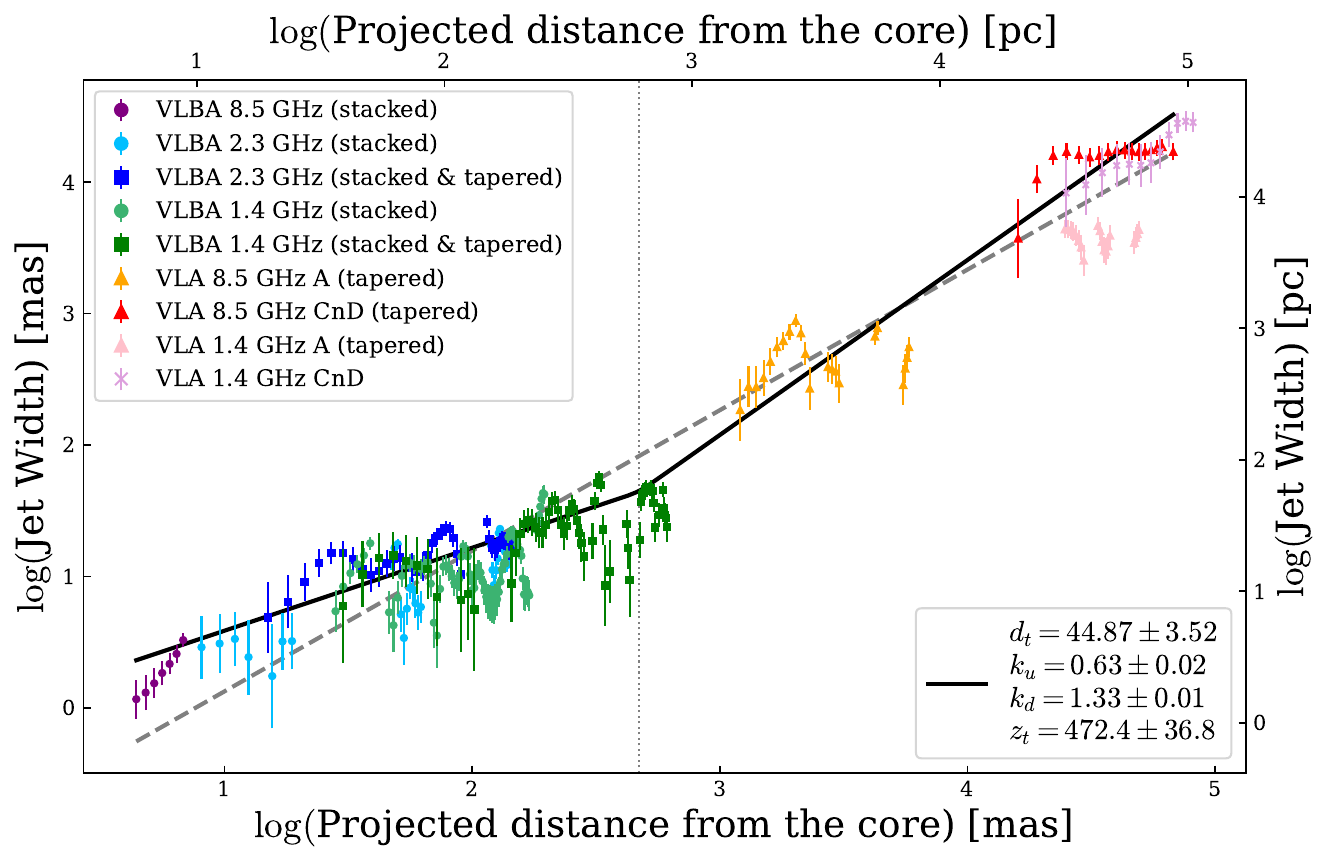}}
\caption{
Jet widths versus projected distances from the core, shown in logarithmic scale.
The black solid line represents a broken power-law fit to the data, using Eq.~\ref{eq:jetwidth}, with a fixed smoothness parameter $h=25$. The dashed gray line corresponds to a single power-law fit to the data.
The width errors include 10\% of the measured value as additional systematic uncertainty, and its minimum is set to one-fifth of the beam size.
}
\label{fig:dist-width}
\end{figure*}

Our main results---the transverse jet width profile as a function of distance from the core---are shown in Fig. \ref{fig:dist-width}, along with the best fit parameters shown as inset. 
The collimation profile exhibits several important features.
First, a single power-law (gray dashed line in Fig.~\ref{fig:dist-width}) clearly fails to describe the collimation behavior over the full spatial range. Instead, a broken power-law model (black solid line in Fig. \ref{fig:dist-width}) provides a significantly better fit. Details of formal calculations of the statistical significances for both models are presented in Appendix \ref{sec:statistics}, and we conclude that broken power-law provides a significantly better description of the observations. In more detail, the inner jet ($z < z_{t}$) follows a power-law slope of $k_{u} = 0.63 \pm 0.02$, broadly consistent with a parabolic collimation profile ($k_{u} \sim 0.5$; e.g., \citealt{Asada_2012}). Beyond the transition point, the slope changes to $k_{d} = 1.33 \pm 0.01$, indicating a slightly over-conical expansion relative to a purely conical jet ($k_{d} = 1$).
Furthermore, the transition occurs at $z_{t} \approx 472$\,mas ($\sim 610$\,pc projected), or $\sim 7.9 \times 10^6 R_{S}$ from the core. This parabolic-to-conical transition is remarkably similar to that observed in various jets from AGNs hosted by elliptical galaxies (e.g., \citealt{Asada_2012,Tseng_2016,Nakahara_2019,Kovalev_2020,Boccardi_2021}).

We further note the following points.
In Fig. \ref{fig:dist-width} we fixed the parameter $h=25$ for the presentation of the result (see also \citealt{Boccardi_2021}). 
While the exact value of $h$ was not strongly constrained by fitting to the data, varying the value of $h$ within reasonable bounds did not change the other model parameters of Eq. \ref{eq:jetwidth} significantly beyond 1$\sigma$ statistical error.
Also, we note that the jet width measurements across different frequencies agree well within uncertainties, indicating that coreshift effects do not introduce significant registration errors in our maps, enhancing the reliability of the collimation profile across the broad spatial range probed.


\section{Discussions \& Conclusions} \label{sec:discussions}

The observed transition from a parabolic to a nearly conical jet shape in 0313--192 strikingly resembles those observed in other AGNs jets (see Sect. \ref{sec:intro}), supporting the idea that this structural evolution may be a generic feature of relativistic jets \citep{kovalev20}, even for AGNs hosted by disk galaxies.
In 0313--192, this transition occurs at a projected distance of $z_{t}\sim610$\,pc$\sim7.9\times10^{6}R_{S}$.
For comparison, we compute the sphere of gravitational influence (SGI) radius $R_{\rm SGI}$ for the SMBH of 0313--192, defined as as $R_{\rm SGI}=GM_{\rm BH}/\sigma_{*}^{2}=(1/2) (c/\sigma_{*})^{2}\,R_{\rm S}$
where $\sigma_{*}$ is the stellar velocity dispersion around the BH.
Using the $M_{\rm BH}-\sigma_{*}$ relation (Eq. 7 of \citealt{Kormendy_2013}), we estimate $\sigma_{*}\sim248$\,km/s and accordingly $R_{\rm SGI}\sim7.3\times10^{5}\,R_{\rm S}$.
This indicates that the jet collimation occurs effectively on scales beyond those directly governed by the gravity of the central SMBH.
In theoretical models, such large-scale jet collimation is often attributed to external pressure confinement by hot interstellar medium (ISM) or intergalactic medium (IGM) with power-law pressure gradients ($p \propto z^{-\kappa}$, $\kappa < 2$), which can maintain parabolic jet shapes and trigger efficient magnetic acceleration \citep{celotti01,komissarov07,komissarov09,lyubarsky09}. In this context, the observed parabolic shape of the inner jet in 0313–192 at $z<z_{t}$ may explain how the jet in this peculiar spiral galaxy could have been successfully collimated, accelerated, and extended up to the large $\sim100$\,kpc scales, producing its DRAGN morphology.

Such scenarios are plausible for elliptical galaxies, where hot halo gas is common due to large-scale cooling flows from the surrounding group or cluster environment and the hot gas can be captured within the Bondi radius to provide the necessary confining pressure \citep{mathews_2003,Yuan_2014}. Indeed, in many elliptical galaxies, the jet collimation break appears near the Bondi radius ($z_{t} \sim 10^5$–$10^6,R_S$; \citealt{Kovalev_2020}).
In contrast, spiral and disk galaxies typically lack extended, pressure-dominated atmospheres. 
Their ISM is instead more clumpy and rotationally supported, making sustained collimation more difficult.
For 0313–192 specifically, neither X-ray nor CO(1–0) observations reveal large-scale gaseous halos around the bulge or cold and dense gas surrounding the SMBH \citep{ledlow2001large,Keel_2006}. 
In Appendix~\ref{sec:calc_pressure}, we further assess alternative ISM components, including a putative circumnuclear disk and stellar wind-driven hot ISM from old stellar population associated with the particularly massive bulge of the galaxy \citep{Keel_2006}. As demonstrated therein, both are found to have thermal pressures many orders of magnitude below the expected pc-scale jet internal pressure.

This leads us to consider non-thermal pressure components or other magnetically mediated collimation mechanisms. 
In particular, several theoretical works show, analytically (e.g., \citealt{bogovalov05,gracia05,beskin06,blandford22}) and also numerically by magnetohydrodynamic (MHD) simulations (e.g., \citealt{Dihingia_2021}), that magnetized winds from accretion disks with relatively wide opening angle (e.g., \citealt{bp82}) can help the jet collimation by ram and magnetic pressure.
Even if those winds may not persist up to the large spatial scales of $z_{t}\sim10^7 R_{S}$ \citep{Yuan_2014,Blandford_2019}, they can still help the initial confinement of the central relativistic jet \citep{bz77} during its early evolution and propagation, influencing internal evolution of the jet from Poynting-flux-dominated to kinetically dominated regimes until the jet reaches the collimation transition point \citep{beskin17}.
We note that such scenarios are also relevant to well collimated jets in other types of objects such as 3C\,273 (quasar; \citealt{Okino_2022}) and 1H\,0323+342 (narrow-line Seyfert 1; \citealt{Hada_2018}), which accrete at high Eddington ratios \citep{Yuan_2014} and may not exhibit Bondi-type inflows \citep{narayan11}.

Yet, this jet collimation scenario still leaves a key open question: why are large-scale relativistic jets so rare in spiral galaxies, if collimation via disk winds is a generally available mechanism? The rarity likely reflects a twofold constraint on jet production and propagation.
As suggested by \cite{Wu_2022}, disk galaxy-hosted DRAGNs tend to have exceptionally massive stellar bulges of $>10^{11}M_{\odot}$ compared to typical late-type galaxies, implying $M_{\rm BH}>10^{8}M_{\odot}$. 
Such massive black holes likely formed through major mergers and prolonged accretion  \citep{volonteri10}, which may result in high black hole spin and the accumulation of poloidal magnetic flux. Both are are necessary for launching powerful black hole jets \citep{tchekhovskoy11}.
Theoretical studies also show that if massive SMBHs grew via coherent accretion in gas-rich environments, which can be relevant for late-type galaxies such as 0313--192 without features of major galaxy mergers, they could attain high black hole spins (e.g., \citealt{bustamante19}), allowing efficient jet production.
We note, however, that this might be an over-simplification, as observations suggest various spin values for massive SMBHs of $M_{\rm BH}\gtrsim10^{8}M_{\odot}$ from modest or low \citep{vasudevan16,reynolds19} to 
high spins \citep{eht19}.
Meanwhile, the production of the disk wind depends on the dynamical state of the accretion disk. Strongly magnetized thin disks may be transient and accretion disks in AGNs hosted by late-type galaxies may not always produce stable winds over long periods but loose their energy by radiation (e.g., \citealt{livio03}), which is supported by wide ranges of radio loudness in quasars hosted by disk galaxies (see \citealt{sikora07} for discussions). 
Thus, the rarity of large-scale relativistic jets in disk galaxies may ultimately be a result of unique combination of high BH spin, large-scale and strong magnetic flux, and a stable accretion disk capable of producing collimating winds,
as other studies of spiral DRAGNs have noted (e.g., \citealt{Bagchi2014}).


In summary, our analysis of the relativistic jet in the spiral galaxy 0313--192 reveals a distinct parabolic-to-conical transition in its collimation profile, with the break occuring at a projected distance of $\sim7.9\times10^6 R_{S}$. This transition, located far beyond the $R_{SGI}\sim7.3\times10^{5} R_{S}$, points to an extended collimation zone, revealing the role of MHD processes as the dominant mechanism shaping jet collimation and propagation in this peculiar system. The lack of sufficient thermal pressure from the galactic ISM, on the other hand, suggests that the observed jet is likely confined by other non-thermal components, such as ram and magnetic pressure from disk winds. 
Our findings provide hints about how AGNs hosted by disk galaxies---under favorable conditions for the black hole and accretion disk---can produce large-scale, well-collimated jets that were traditionally thought to be exclusively associated with large elliptical hosts. This emphasizes the need to broaden models of AGN jet formation beyond the classical paradigm of hot, thick-disk accretion in massive elliptical galaxies, and to account for a more diverse range of host environments and accretion state. 
Future higher-resolution and multi-frequency imaging of jet-launching regions (e.g., \citealt{kim18b,kim23}), along with sensitive polarimetry of jet, will be essential for probing the fine-scale structures of the jet base as well as magnetic field properties and structures in similar systems.


\vspace{\baselineskip}

{\sffamily\textit{Acknowledgments}}
The authors thank the anonymous referee for constructive comments and suggestions, which substantially improved the manuscript.
S.Y.L. and J.-Y.K. acknowledge supports for this research provided by
the Dongil Culture and Scholarship Foundation 
and
the National Research Foundation of Korea (NRF) grant funded by the Korean government (Ministry of Science and ICT; grant no. 2022R1C1C1005255, RS-2022-00197685 and RS-2024-00396244).
The National Radio Astronomy Observatory is a facility of the National Science Foundation operated under cooperative agreement by Associated Universities, Inc.
This work made use of the Swinburne University of Technology software correlator \citep{deller11}, developed as part of the Australian Major National Research Facilities Programme and operated under licence. 
This research has made use of NASA's Astrophysics Data System Bibliographic Services.
This work made use of Astropy\footnote{http://www.astropy.org}: a community-developed core Python package and an ecosystem of tools and resources for astronomy.


\nocite{*}
\bibliography{0313-192_ref}


\appendix

\section{Additional Table and Figures}\label{sec:more_details}

\begin{figure*}[htbp]
\centering
\gridline{%
  \fig{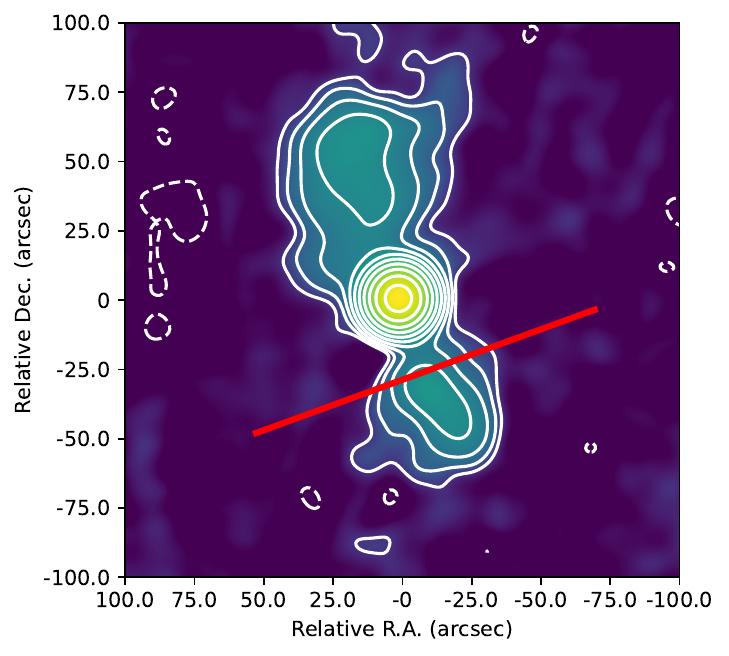}{0.42\textwidth}{}%
  \fig{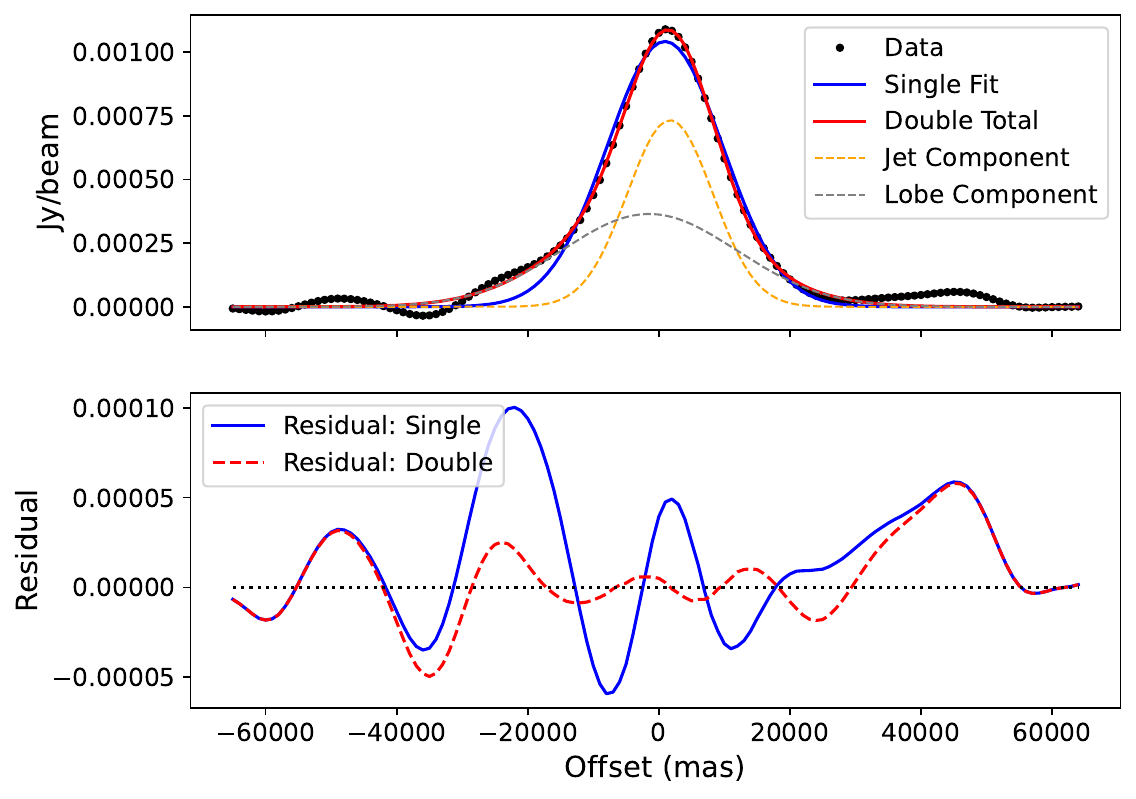}{0.56\textwidth}{}%
}
\caption{%
Example of double-Gaussian fitting to a cross-jet intensity slice profile, in order to decompose the jet and lobe emission. 
Left: an example VLA X-band, $(u,v)$-tapered CnD configuration image of 0313--192 (see Sect. \ref{sec:results}). The red line shows a slice used to make the intensity profile on the right panel. 
Right: Example of double Gaussian fitting to the jet slice intensity profile. The data, fitted profiles for the single and double-Gaussian cases, and the individual jet and lobe components are shown in the upper panel. The corresponding fit residuals are shown in the lower panel.
}
\label{fig:XCnD_slicing}
\end{figure*}

\begin{figure*}[h!]
\centering
\resizebox{0.85\textwidth}
{!}{\plotone{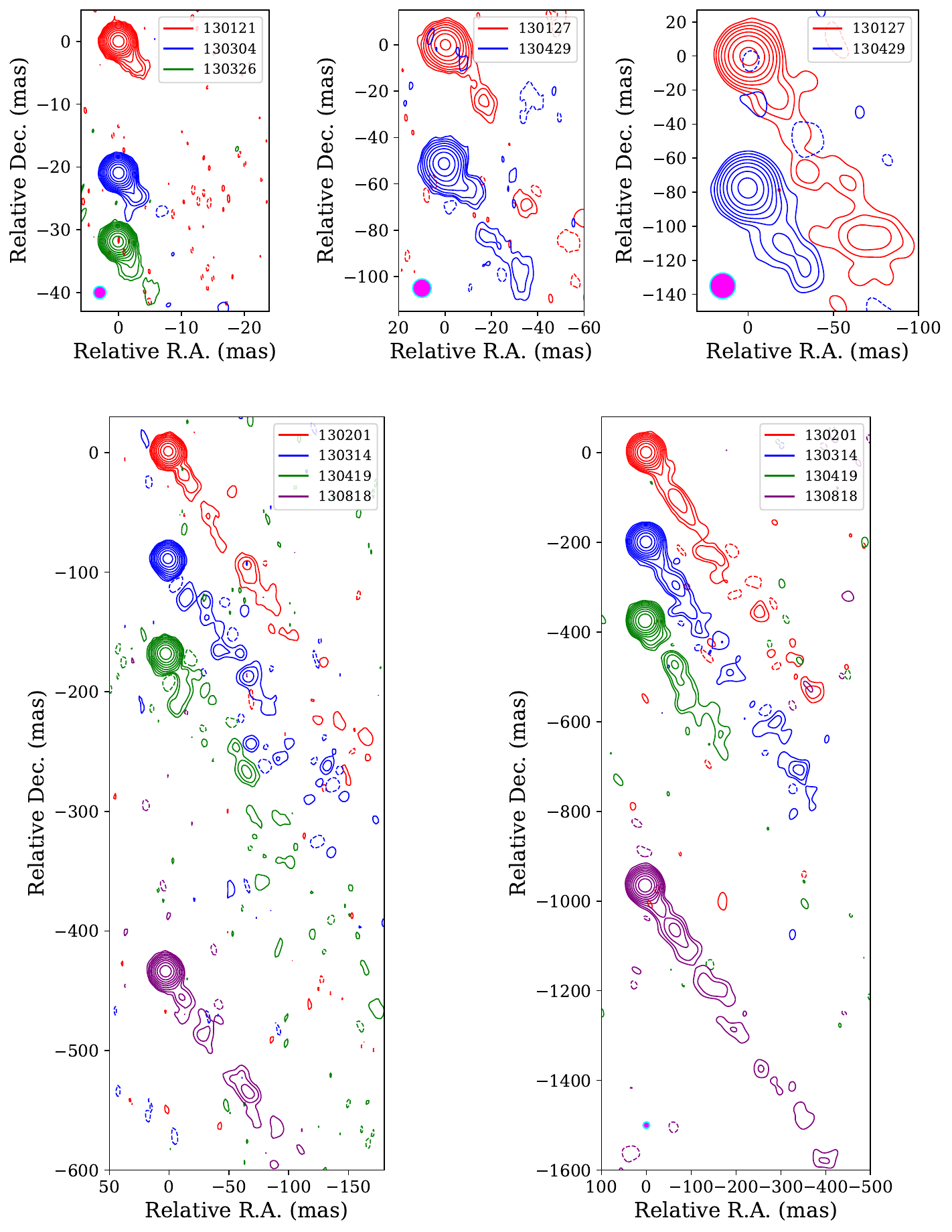}}
\caption{
Individual VLBA images of 0313–192 across different observing bands, tapering, and epochs, arranged vertically by observing date (in yymmdd format). 
From top left to bottom right, panels show the VLBA images at increasing beam sizes: raw-resolution X-band (top-left), the raw and $(u,v)$-tapered S-band (top-center and top-right), and the raw and tapered L-band images (bottom row). 
Within each panel, the restoring beam sizes (2, 8, 15, 10, and 30 milliarcseconds, respectively) are shown by magenta circles located at the lower-left corner of each panel.
Also, separation between images is proportional to the time difference of the observing epochs.
In all images, contour levels start at 2.5\,$\sigma$ and increase by factors of 2.
}
\label{fig:XSL-imaging1}
\end{figure*}

Here we present additional Table and Figures.
Table \ref{tab:data} shows details of the observational data and images used for this study.
Figure \ref{fig:XCnD_slicing} shows example of double-Gaussian fitting of a slice intensity profile applied for the VLA X-band CnD configuration image.
Figure \ref{fig:XSL-imaging1} displays all the individual images of 0313--192 obtained by the VLBA using various weighting and tapering.

\begin{deluxetable*}{cccccccccc}[h!]
\caption{Summary of the observations and image parameters.}
\label{tab:data}
\tablehead{
\colhead{Array} & 
\colhead{Band} & 
\colhead{$\nu$ (GHz)} & 
\colhead{Obs ID} & 
\colhead{Epoch} &
\colhead{$uv$-taper} & 
\colhead{$uv$-weight} &
\colhead{beam (mas)} &
\colhead{peak (mJy/beam)} &
\colhead{rms ($\mu$Jy/beam)} \\
\colhead{(1)} &
\colhead{(2)} &
\colhead{(3)} &
\colhead{(4)} &
\colhead{(5)} &
\colhead{(6)} &
\colhead{(7)} &
\colhead{(8)} &
\colhead{(9)} &
\colhead{(10)}
}
\startdata
VLA & X & 8.5 & AL400CnD & 971012 & 0, 0 & 0, $-1$ & 5000.0 & $73.5\pm7$ & $31$ \\
VLA & X & 8.5 & AL400CnD & 971012 & 0.1, 0.02 & 0, $-1$ & 13000.0 & $74.3\pm7$ & $40$ \\
VLA & X & 8.5 & AY119A & 001022 & 0.1, 1 & 0, $-1$ & 500.0 & $75.1\pm8$ & $62$ \\
VLA & L & 1.4 & AL400CnD & 971012 & 0, 0 & 5, 0 & 25000.0 & $50.4\pm5$ & $225$ \\
VLA & L & 1.4 & AY119A & 001022 & 0, 0 & 0, $-1$ & 2500.0 & $50.4\pm5$ & $225$ \\
VLA & L & 1.4 & AY119A & 001022 & 0.1, 0.1 & 0, $-1$ & 3500.0 & $34.0\pm3$ & $181$ \\
VLBA & X & 8.5 & BM376A & 130121 & 0, 0 & 0, $-1$ & 2.0 & $65.4\pm7$ & $32$ \\
VLBA & X & 8.5 & BM376B & 130304 & 0, 0 & 0, $-1$ & 2.0 & $69.3\pm7$ & $22$ \\
VLBA & X & 8.5 & BM376C & 130321 & 0, 0 & 0, $-1$ & 2.0 & $66.3\pm6$ & $22$ \\
VLBA & S & 2.3 & BM376D & 130127 & 0, 0 & 0, $-1$ & 8.0 & $27.5\pm3$ & $84$ \\
VLBA & S & 2.3 & BM376D & 130127 & 0.1, 20 & 0, $-1$ & 15.0 & $32.4\pm3$ & $62$ \\
VLBA & S & 2.3 & BM376E & 130426 & 0, 0 & 0, $-1$ & 8.0 & $25.9\pm3$ & $90$ \\
VLBA & S & 2.3 & BM376E & 130426 & 0.1, 20 & 0, $-1$ & 15.0 & $29.6\pm3$ & $106$ \\
VLBA & S & 2.3 & BM376F & 130429 & 0, 0 & 0, $-1$ & 8.0 & $31.2\pm3$ & $73$ \\
VLBA & S & 2.3 & BM376F & 130429 & 0.1, 20 & 0, $-1$ & 15.0 & $34.3\pm3$ & $73$ \\
VLBA & L & 1.4 & BM376G & 130201 & 0, 0 & 0, $-1$ & 10.0 & $28.7\pm3$ & $25$ \\
VLBA & L & 1.4 & BM376G & 130201 & 0.1, 10 & 0, $-1$ & 30.0 & $31.1\pm3$ & $28$ \\
VLBA & L & 1.4 & BM376H & 130314 & 0, 0 & 0, $-1$ & 10.0 & $29.9\pm3$ & $29$ \\
VLBA & L & 1.4 & BM376H & 130314 & 0.1, 10 & 0, $-1$ & 30.0 & $32.2\pm3$ & $24$ \\
VLBA & L & 1.4 & BM376H2 & 130818 & 0, 0 & 0, $-1$ & 10.0 & $35.5\pm4$ & $30$ \\
VLBA & L & 1.4 & BM376H2 & 130818 & 0.1, 10 & 0, $-1$ & 30.0 & $38.0\pm3$ & $30$ \\
VLBA & L & 1.4 & BM376I & 130419 & 0, 0 & 0, $-1$ & 10.0 & $34.7\pm3$ & $23$ \\
VLBA & L & 1.4 & BM376I & 130419 & 0.1, 10 & 0, $-1$ & 30.0 & $36.9\pm4$ & $25$ \\
VLBA & $X^*$ & 8.5 & $-$ & $-$ & 0, 0 & 0, $-1$ & 2.0 & $68.4\pm7$ & $15$ \\
VLBA & $S^*$ & 2.3 & $-$ & $-$ & 0, 0 & 0, $-1$ & 8.0 & $28.9\pm3$ & $47$ \\
VLBA & $S^*$ & 2.3 & $-$ & $-$ & 0.1, 20 & 0, $-1$ & 15.0 & $32.1\pm3$ & $46$ \\
VLBA & $L^*$ & 1.4 & $-$ & $-$ & 0, 0 & 0, $-1$ & 10.0 & $32.2\pm3$ & $13$ \\
VLBA & $L^*$ & 1.4 & $-$ & $-$ & 0.1, 10 & 0, $-1$ & 30.0 & $34.6\pm3$ & $14$ \\
\enddata
\tablenotetext{}{
\textbf{Note.} 
(1) Observing arrays (either VLA or VLBA). 
(2) Observing bands. 
(3) Central observing frequency in GHz. 
(4) NRAO project code. 
(5) Observing dates in yymmdd format. 
(6) and (7) Parameters for the \texttt{uvtaper} (weight amplitude, $(u,v)$-radius) and \texttt{uvweight} (weighting bin width, error power) commands in Difmap, respectively.
(8) FWHM sizes of circular beams in mas. 
(9) and (10) Peak flux density and rms noise in mJy/beam and $\mu$Jy, respectively. As for the peak flux density, 10\% systematic error is assumed \citep{mao2018}. 
\tablenotetext{*}{Final images made by stacking the corresponding VLBA band maps and used for the jet width measurement.}
}
\vspace{-25pt}
\end{deluxetable*}


\section{Statistical significance of the broken power-law fit}\label{sec:statistics}

Here we discuss statistical significance of the broken power-law model for the jet width over the single power-law fit.
Besides the broken power-law (Eq. \ref{eq:jetwidth}), we also fitted following single power-law model to the jet width measurements: 
\begin{equation}
    d(z)=d_{0}(z/z_{0})^{k}
    \label{eq:single}
\end{equation}
where $d_{0}$ is the jet width at a chosen distance $z_{0}$ and $k$ is the power-law slope of the width profile (i.e., thus two free parameters).
To formally estimate statistical significance of these models, we computed the Bayesian information criterion (BIC) \citep{liddle07} for the Gaussian likelihood case, as
\begin{equation}
    BIC=\chi^{2}+p\ln(N)
    \label{eq:bic}
\end{equation}
where $\chi^{2}=\sum_{i}( (D_{i}-M_{i})^{2}/\sigma^{2}_{i} )$ is the total chi-square computed from the data and model, $D$ and $M$ are the data and model prediction, $\sigma$ is the uncertainty of the data, $p$ is the number of free parameters used in the model, and $N$ is the number of data points used for the fitting.
For the single power-law model, we obtain the total $\chi^{2}\sim12000$ (reduced $\chi^{2}\sim73$), while for the broken power-law model we find a much smaller $\chi^{2}\sim14700$ (reduced $\chi^{2}\sim88$).
Given $p=2$ and $6$ for the single and broken power-law models, we find relative excess of the BIC for the single power-law model of $\Delta$BIC$\sim2700$. This difference is substantial to discard the single power-law model \citep{liddle07}.

\section{Thermal pressure due to various ISM components in the galaxy}\label{sec:calc_pressure}

Here we estimate the thermal pressure contributed by different ISM components in the inner pc-scale region of 0313--192, and compare these values to the expected jet internal pressure.
Our first calculation involves a tilted, arcsecond-scale circumnuclear emission-line structure reported by \citet{Keel_2006}, which may represent either a gas disk formed through a minor galaxy merger or an ionization cone driven by the AGN. We assume it corresponds to a gaseous disk potentially linked to a warm, dense absorber near the nucleus, based on the unusually strong HI and X-ray absorption (column density $N_{\rm H} \sim 4 \times 10^{22}$\,cm$^{-2}$) compared to typical edge-on Seyfert nuclei \citep{ledlow2001large,Keel_2006}. Assuming a line-of-sight depth of $r \sim 100$\,pc and a temperature of $T \sim 600$\,K \citep{Keel_2006}, the thermal pressure is
\[
P_{\rm th} = nk_{\rm B}T = \left(\frac{N_{\rm H}}{r}\right)k_{\rm B}T \sim 10^{-11}~\mathrm{dyne~cm^{-2}},
\]
\noindent
where $n$ is the particle number density. This value is higher than the total jet internal pressure on kpc scales ($P_{\rm jet} \sim 10^{-13}$\,dyne\,cm$^{-2}$; \citealt{ledlow2001large}), but lower than the expected pc-scale jet pressure by several orders of magnitude:
\[
P_{\rm jet} > P_{B} = \frac{B^2}{8\pi} \sim 10^{-5}\text{--}10^{-4}~\mathrm{dyne~cm^{-2}}
\]
\noindent
where $P_B$ is the magnetic pressure and $B \sim 0.01-0.1$\,G is the assumed equipartition magnetic field strength on pc scales (somewhat lower than the typical $B_{\rm eq} \sim 0.4-0.9$\,G for pc-scale jets in quasars and BL Lacs; \citealt{Pushkarev_2012}). Even if we allow for uncertainties by factors of a few in path length or field strength, the pressure mismatch by a factor of $\sim 10^5-10^6$ suggests that this warm disk is unlikely to be the main collimating medium for the pc-scale jet.

Our second estimate considers hot stellar winds from the old stellar population in the bulge. The unusually bright bulge of 0313--192 (implying large stellar mass of $M_* \sim 3 \times 10^{11}\,M_\odot$; \citealt{Keel_2006}) and its large effective radius ($\sim 2.5'' \sim 4 \times 10^7\,R_{\rm S}$) suggest probable abundant supply of stellar wind. Assuming 10\% of the bulge stars are in the red giant branch (RGB) phase and actively losing mass, we adopt the following stellar wind parameters: 
mass-loss rate per star
\[
\dot{M} \sim 10^{-13}\,M_\odot\,\left(\frac{L}{L_\odot}\right)\left(\frac{R}{R_\odot}\right)\left(\frac{M}{M_\odot}\right)^{-1}\,\mathrm{yr^{-1}},
\]
\citep{reimers75,schroeder05}
\noindent
with typical values $L = 10^3\,L_\odot$, $R = 100\,R_\odot$, $M = 1\,M_\odot$, wind velocity $v \sim 300$\,km\,s$^{-1}$, and temperature $T = 10^5$\,K. Assuming spherical inflow through a radius of $r = 10$\,pc, the total inflow rate is $\dot{M}_{\rm in} = 0.1 \times (3 \times 10^{11}) \dot{M}$ and the resulting mass flux density at the 10\,pc is
\[
\rho = \frac{\dot{M}_{\rm in}}{4\pi r^2 v}.
\]
Assuming fully ionized gas ($\mu = 0.5$), the particle number density is $n = \rho / \mu m_{\rm p}$, leading to a thermal pressure
\[
P_{\rm th} = nk_{\rm B}T \sim 9 \times 10^{-7}~\mathrm{dyne~cm^{-2}}.
\]
This pressure is significantly higher than that from the putative circumnuclear disk, but still is significantly smaller than the expected jet pressure by a factor of $\sim 10-100$.

Thus, unless currently available multi-wavelength observations of 0313--192 \citep{Keel_2006} have missed other significant ISM components due to sensitivity limits, we can conclude that the galactic environment alone is insufficient to provide the pressure needed to collimate the inner jet of 0313--192.


\end{document}